\documentclass[apj]{jemulateapj}

\usepackage{natbib,enumerate,psfig}
\usepackage{graphicx}

\shorttitle{Johnson \& Wright respond to Lloyd 2013}
\shortauthors{Johnson \& Wright}

\submitted{Informal arXiv submission}
\begin{document}

\title{On Lloyd's {\it The Mass Distribution of Subgiant Planet Hosts} (\lowercase{ar}X\lowercase{iv:1306.6627v1})}

\author{John Asher Johnson}
\affil{Department of Astronomy, California Institute of Technology, 1200 E. California Blvd., Pasadena, CA 91125, USA\\ 
NASA Exoplanet Science Institute (NExScI), CIT Mail Code 100-22, 779 South Wilson Avenue, Pasadena, CA 91125, USA\\johnasherjohnson@gmail.com}

\and
\author{Jason T.\ Wright}
\affil{Department of Astronomy and Astrophysics, 525 Davey Lab, The Pennsylvania State University, University Park, PA 16802, USA \\
Center for Exoplanets and Habitable Worlds, 525 Davey Lab, The Pennsylvania State University, University Park, PA 16802, USA \\jtwright@astro.psu.edu}

\begin{abstract}

We provide an informal response to James P.\ Lloyd's recent arXiv
preprint (arXiv:1306.6627v1) ``The Mass Distribution of Subgiant
Planet Hosts'' accepted for publication by {\it Astrophysical Journal
  Letters}.  

\end{abstract}

\section{Context}

Recently, James P. Lloyd posted a preprint on the arXiv
\citep[arXiv:1306.6627v1,][]{Lloyd(2013)} of {\it The Mass Distribution of Subgiant Planet
Hosts} accepted for publication by Astrophysical Journal Letters.
Lloyd's preprint is a surrebuttal to our paper {\it Retired A Stars: The
Effect of Stellar Evolution on the Mass Estimates of Subgiants}
\citet[][hereafter JMW13]{Johnson et al.(2013)}, which in turn is a rebuttal to
Lloyd's manuscript {\it ``Retired'' Planet Hosts: Not So Massive, Maybe Just
Portly After Lunch} \citep[][hereafter L11]{Lloyd(2011)}.

Exploring (and, hopefully, settling) this issue of the true masses of
subgiant stars deserves careful analysis and peer review, and this is
best undertaken in a thorough discussion in the refereed
literature. In addition to Lloyd's work, the recent work by \citet{Schlaufman
and Winn(2013)}, suggesting that the kinematics of the
subgiant planet hosts are inconsistent with having evolved from an
A-star main sequence population, deserves close attention.  We are
undertaking observations and preparing work for future publication on
these topics, including interferometric radii and asteroseismological
masses of bright subgiants. 

That said, since Lloyd's paper has generated immediate
interest, as well as calls for us to quickly respond, we are
making this post to the arXiv to share our considered thoughts on it,
for the benefit of the astronomical community. 

To be clear, the comments below do not speak directly to the
resolution of the many outstanding questions regarding of the true
masses of subgiant planet host stars, but pertain specifically to our
thoughts on Lloyd's recent paper {\it per se}. 

\section{Response}

In L11, Lloyd argued that the spectroscopically derived masses from
our subgiant planet search were systematically high, largely because
the number of “retired A stars” in our sample appeared inconsistent
with his models of stellar evolution. L11 attributed this difference
to shape of the IMF and the differing rates of evolution across the
subgiant branch for stars of different masses. Both effects should
reduce the number of observable, massive subgiants. This inspired us
to check our spectroscopically-derived masses for consistency with
expectations from state-of-the-art galactic population synthesis
models, which we described in JMW13. 

We showed results from one such model, TRILEGAL \citep{Girardi et al.(2005)},
which accurately reproduces the Hipparcos color-magnitude diagram, as
well as star counts from many different surveys such as 2MASS. Figure
1 shows a simulated color-magnitude diagram of a 200 pc
sample of stars (red) compared to the Hipparcos sample (green)
(L. Girardi private communication). In JMW13, we selected stars from
the subgiant branch of this simulated sample using the color and
apparent/absolute magnitude cuts described by Johnson et al. The
resulting distribution of stellar masses agrees well with the stellar
masses in the planet search sample of \citet{Johnson et al.(2010)} and allows
for a fraction of those subgiants to have masses in excess of 1.5
Msun. 

\begin{figure*}
\plotone{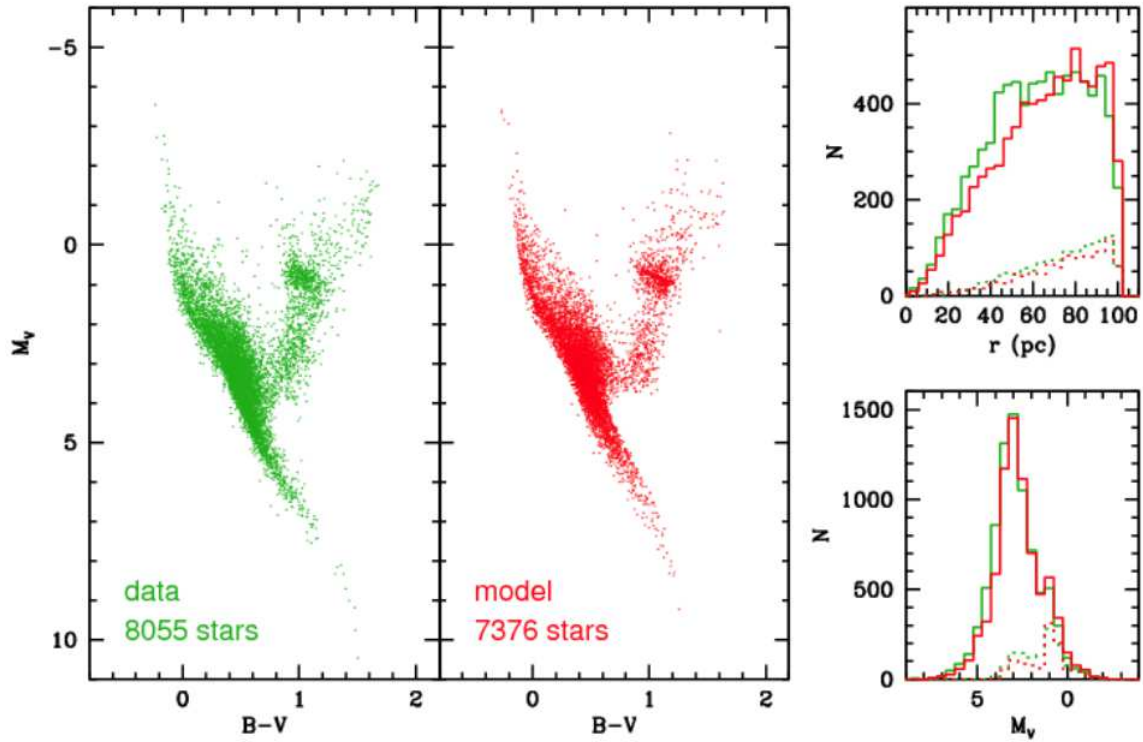}
\caption{Comparison of TRILEGAL galactic synthesis to data from {\it
    Hipparcos} in color-magnitude, distance, and absolute magnitude
  spaces (L.\ Girardi, private communication).  TRILEGAL incorporates all of
the physics Lloyd invokes to explain why subgiants with the masses we
ascribe to them in our previous works should be very rare, but
TRILEGAL predicts that they should be in our sample, nonetheless.
Despite the good match between TRILEGAL's model outputs and
observations, Lloyd prefers his models, which have not been tested in
this manner and which give different results.   } 
\end{figure*}

Given this, we feel that there is now a significant burden on Lloyd to
demonstrate not just that spectroscopically-derived masses are
problematic, but to directly address the demonstrations in JMW13 and
argue why we should not trust the TRILEGAL simulations. He has not done
this in his newest manuscript. Instead, he uses his own galactic population
model, the details of which have not been described in the refereed
literature nor tested on actual data.  

Lloyd uses his galactic population model to argue, contra JMW13, that
the Malmquist bias is not the real difference for the mass
distributions of L11 and JMW13.  Instead, he argues that the
difference among stars with M$<1.3$ solar is actually a difference in
the scale heights chosen by us and him. This is possible, indeed
perhaps likely, but Lloyd does not actually demonstrate this to be the
case.  

More problematically, Lloyd does not argue that the TRILEGAL scale
height is wrong, just that the answer is sensitive to it.  Given the
success of TRILEGAL in reproducing the Hipparcos color-magnitude
diagram and 2MASS star counts (among other stellar catalogs), we feel
that this argument implicates Lloyd's models, not TRILEGAL's, as
having the wrong scale height.   

We note that Lloyd could address both of these issues using TRILEGAL
itself by changing the scale height parameter and rerunning our
simulations.  We have provided Lloyd with the scripts we used to call
TRILEGAL, so he could easily make a direct comparison with JMW13; with
these he can also easily check whether a different scale height
reproduces the L11 results, and whether it is consistent with the
Hipparcos star counts and color-magnitude diagram. 

In Section 2.4, Lloyd argues heuristically why high-mass subgiants are
rare, but all of the physics he invokes are already accounted for in
TRILEGAL, so this does not actually support his contention that our
mass distribution is wrong. He addresses the analytic scaling
arguments we used in JMW13, but our conclusions never rested on those
simple calculations and JMW13 is clear on this point. Again, the
burden is on Lloyd to show not only that the output of TRILEGAL is
sensitive to certain model parameters (which we do not dispute) but
that the star formation rates and the age-metallicity relations used
by TRILEGAL are wrong, which he does not do. 

Why should the reader trust the L11 models over TRILEGAL? Lloyd's sole
argument appears to be that the planet-occurrence rate as a function
of mass implied by our mass estimates has an “implausible” shape. 

\citet{Johnson et al.(2010)} already performed a robust calculation of the
mass dependence of this rate, based on both the detections and
non-detections of the planet surveys at Keck and Lick
observatories. Lloyd is apparently attempting to re-derive the
occurrence rate calculations of \citet{Johnson et al.(2010)} by dividing the
cumulative mass distribution function of planet-hosting stars reported
in the literature --- as detected from a heterogeneous set of surveys
with a variety of stellar selection criteria --- by the mass CDF of
stars expected to be in such a region of color-magnitude space from
various population synthesis models. 

This does not appear to be a robust way to determine the mass dependence of planet
frequency, and Lloyd provides no error analysis to demonstrate its
precision or accuracy.  We suggest that the implausible nature of the resulting planet frequency function with
mass is a result of the rough nature of the planet frequency analysis,
not the underlying mass estimates.  

Lloyd is attempting to point out large systematic errors in spectral
synthesis models (SME), stellar evolution models, galactic population
models and planet occurrence rate calculations.  These claims require
more details than are offered in this paper, and at any rate many of
the points in the paper do not seem to actually support its thesis.

\acknowledgments

We acknowledge our co-author Tim Morton for his significant
contribution both to our work on the subject and this document.

We would also like to use this informal forum to note Dr. Lloyd's
transparent and professional approach to this controversy, including
his solicitation of our input before submission and frank discussions
on other occasions.  He has repeatedly identified ways to make our
mass estimates and manuscripts better, and we thank him for his care
and dedication to this important topic. Although our disagreement on
this scientific topic is profound, the dispute is not personal, and we
remain friends and colleagues.


\begin{thebibliography}{}
\bibitem[Girardi et al.\ (2005)]{Girardi et al.(2005)} Girardi, L.,
  Groenewegen, M.~A.~T., Hatziminaoglou, E., \& da Costa, L.\ 2005, \aap, 436, 895 
\bibitem[Johnson et al.\ (2010)]{Johnson et al.(2010)} Johnson, J.~A.,
  Aller, K.~M., Howard, A.~W., \& Crepp, J.~R.\ 2010, \pasp, 122, 905 
\bibitem[Johnson et al.\ (2013)]{Johnson et al.(2013)} Johnson, J.~A., Morton, 
T.~D., \& Wright, J.~T.\ 2013, \apj, 763, 53 
\bibitem[Lloyd (2011)]{Lloyd(2011)} Lloyd, J.~P.\ 2011, \apjl, 739, L49 
\bibitem[Lloyd (2013)]{Lloyd(2013)} Lloyd, J.~P.\ 2013, arXiv:1306.6627 
\bibitem[Schlaufman \& Winn(2013)]{Schlaufman and Winn(2013)} Schlaufman, K.~C., \& Winn, J.~N.\ 2013, arXiv:1306.0567 



\end{thebibliography}
\end{document}